\documentclass[showpacs,twocolumn,superscriptaddress,nofootinbib]{revtex4}

\usepackage{doi}
\usepackage{hyperref}
\hypersetup{
  colorlinks=true,        
  linkcolor=blue,         
  citecolor=cyan,         
}

\usepackage{graphicx}
\usepackage{dcolumn}
\usepackage{bm}
\usepackage{color}
\usepackage{footmisc}

\usepackage{amsmath}
\usepackage{amssymb}


\begin{document}

\title{Five dimensional charged rotating minimally gauged supergravity black hole cannot be over-spun and/or over-charged in non-linear accretion}

\author{Sanjar Shaymatov}
\email{sanjar@astrin.uz}

\affiliation{Ulugh Beg Astronomical Institute, Astronomicheskaya 33, Tashkent 100052, Uzbekistan}
\affiliation{National University of Uzbekistan, Tashkent 100174, Uzbekistan}
\affiliation{Tashkent Institute of Irrigation and Agricultural Mechanization Engineers,\\ Kori Niyoziy 39, Tashkent 100000, Uzbekistan}

\author{Naresh Dadhich} 
\email{nkd@iucaa.in}

\affiliation{Inter University Centre for Astronomy \& Astrophysics, Post Bag 4, Pune 411007, India }

\author{Bobomurat Ahmedov} 
\email{ahmedov@astrin.uz}

\affiliation{Ulugh Beg Astronomical Institute, Astronomicheskaya 33, Tashkent 100052, Uzbekistan}
\affiliation{National University of Uzbekistan, Tashkent 100174, Uzbekistan} 
\affiliation{Tashkent Institute of Irrigation and Agricultural Mechanization Engineers,\\ Kori Niyoziy 39, Tashkent 100000, Uzbekistan}

\author{Mubasher Jamil}
\email{mjamil@zjut.edu.cn}

\affiliation{Institute for Theoretical Physics and Cosmology, Zheijiang University of Technology, Hangzhou, China}
\affiliation{Department of Mathematics, School of Natural Sciences (SNS), National
University of Sciences and Technology (NUST), H-12, Islamabad, Pakistan}

\date{\today}
\begin{abstract} 
Generally black hole could be over charged/spun violating the weak cosmic censorship conjecture (WCCC) for linear order accretion while the same is always restored back for non-linear accretion. The only exception however is that of a five dimensional rotating black hole with single rotation that cannot be overspun even at linear order. In this paper we investigate this question for a five dimensional charged rotating minimally gauged supergravity black hole and show that it could not be overspun under non-linear accretion and thereby respecting WCCC. However in the case of single rotation WCCC is however also respected for linear accretion when angular momentum of accreting particle is greater than its charge irrespective of relative dominance of charge and rotation parameters of the black hole.

\end{abstract}
 \maketitle

\section{Introduction}
\label{introduction}

Black holes {have always been very exciting and interesting objects both for their amazing gravitational as well as geometrical properties,  but they have now taken the center-stage after the discovery of gravitational waves produced by merger of two stellar mass black holes in the LIGO-VIRGO detection experiment \cite{Abbott16a,Abbott16b}.} In the near future it is envisaged that gravitational wave observations may uncover some of the hidden properties of black holes which were otherwise not accessible. One of the most fundamental questions in general relativity (GR) is of course testing of the cosmic censorship conjecture (CCC) which has so far remained unproven \cite{Penrose69}. The physical possibility of its violation in the weak form (WCCC) has of late been a very active area of research\footnote{Weak cosmic censorship conjecture essentially states that central singularity is always hidden behind an event horizon and hence is never visible to outside observer \cite{Penrose69,Wald97} under test particle/field accretion.}.

A gedanken experiment {was envisaged in which  over-charged/rotating test particles were bombarded into a black hole to see whether an extremal black hole could be turned into extremal black hole \cite{Wald74b}? The answer turned out to be negative and it was shown that particles with over extrtemal parameters cannot reach horizon of extremal black hole and thereby horizon cannot be destroyed. Thus extremal black hole obeys WCCC under linear test particle accretion. On the other hand it was also shown that a non-extremal black hole can never be turned into extremal \cite{Dadhich97} because as extremality is approached, the allowed window of parameter space of particles with appropriate parameters to reach the horizon pinches off. Thus extremality or zero black hole temperature can never be attained. However the interest in this question got revived when it was argued that a non-extremal black hole cannot be converted into extremal and subsequently extremal to over extremal but extremality could be jumped over to create over extremal state. That is, a black hole could be overcharged \cite{Hubeny99} or overspun \cite{Jacobson09} by a discrete discontinuous accretion process. Thus a naked singularity could be created defying WCCC. On the other hand, a naked singularity was also addressed with a different prospective that whether it could be created as an end state of gravitational collapse~\cite{Joshi93,Joshi00,Goswami06,Giacomazzo-Rezzolla11,Stuchlik14,Joshi15}

{This led to a spurt in activity where various authors studied overcharging/spinning of black holes in different settings violating WCCC, \cite[see,e.g.][]{Saa11,Bouhmadi-Lopez10,Li13,Rocha14,Shaymatov15,Gwak16,Natario16,Song18,
Duztas18,Jana18,Duztas-Jamil18a,Duztas-Jamil18b,Shaymatov19b}}. In all these works, it was assumed that test particle follows a geodesic (or Lorentz force when charged) motion and back and radiation reaction as well as self force effects were not included. {It is though expected  that when these effects will be taken into account, there would be no overcharging/spinning and destruction of black hole horizon \cite{Barausse10,Rocha11,Isoyama11,Zimmerman13,Colleoni15a,Colleoni15b}. {Recently charged scalar and test fields have also been considered for testing WCCC~\cite{Gwak19,Natario20}}. What happens is that particles/fields that could cause over extremal state would not be able to reach black hole horizon. This was precisely how extremality was not destroyed or attained \cite{Wald74b,Dadhich97}. Note that in test particle accretion black hole is perturbed linearly while realistic accretion process like fluid flow would involve non-linear perturbations which could alter the situation completely. This is what has recently been done.} 

{An extensive analysis of non-linear accretion/perturbations has been carried out in a breakthrough work \cite{Sorce-Wald17} leading to the expected result that black hole horizon cannot indeed be destroyed and thereby reestablishing validation of WCCC. The same conclusion was also obtained for Kerr-AdS black hole \cite{Gwak18a}.  Following \cite{Sorce-Wald17}, a number of works have been done of non-linear perturbations {\cite{Ge18,Ning19,Wang19,Yan-Li19,Jiang19}} reinforcing the result that black hole cannot be over charged/spun and horizon cannot be destroyed. Further the same analysis has been done in higher dimension \cite{An18} as well, showing that five dimensional Myers-Perry rotating black hole  \cite{Myers-Perry86} though could be overspun at linear order but when second order perturbations are taken into account the situation reverses --- no overspinning is allowed  and  WCCC is restored. In this case there is yet another subtler case of a black hole with single rotation that cannot be overspun even at linear order, however like all other cases it could however be overspun when both rotations are present~\cite{Shaymatov19a}. However, the six-dimensional rotating black hole with two rotations cannot be
overspun under linear order perturbation~\cite{Shaymatov20a}. A charged black hole in higher dimensions could always be overcharged at linear order \cite{Revelar-Vega17}.}      

{In this paper we would like to examine this question of linear and non-linear accretion for a charged rotating black hole in five dimension. In four dimension, it was straight forward to add charge parameter in the $\Delta$ function of rotating solution;i.e. $\Delta = r^2 - 2Mr +a^2 +Q^2$. Unfortunately this does not work in five dimension, and in fact an analogue of Kerr-Newman black hole has not yet been found. There exists a solution in slow rotation limit \cite{Aliev06,Aliev07a,Aliev07}, and some solutions in supergravity and string theory~ \cite{Cvetic96a,Youm99a,Cvetic96b,Cvetic96c,Cvetic04a, Cvetic04b,Chong05a}. The closest that comes to Kerr-Newman black hole is the one describing minimally gauged supergravity black hole \cite{Chong05}. Black hole energetics  in terms of ergosphere and energy extraction of this solution has been investigated \cite{Prabhu10}. We shall take this solution (by setting $\Lambda = 0$) of minimally gauged supergravity black hole for a charged and rotating black hole in five dimension and study linear and non-linear accretion for testing WCCC. }

{In particular it would be interesting to examine the case of single rotation for linear accretion where black hole cannot be overspun \cite{Shaymatov19a} but could be overcharged \cite{Revelar-Vega17}. It turns out that the ultimate behavior would be determined by relative dominance of angular momentum and charge of accreting particle. If the former is dominant, black hole cannot be over extremalized while if it is the latter, it could be. }

{The paper is organized as follows: In Secs~\ref{Sec:five} and \ref{Sec:Perturbation}, we describe the black hole metric and its properties and build up background for studying linear and non-linear accretion for over extremalizing black hole in the Sec.~\ref{Sec:extremal}. Finally we conclude with a discussion in the Sec.~\ref{Sec:Conclusion}. We shall use the natural units, $G=c=1$ throughout.}

\section{The black hole metric and its properties }\label{Sec:five}

The metric of five dimensional charged and rotating black hole in minimally gauged supergravity black hole \cite{Chong05} is given in the Boyer-Lindquist coordinates $(t,r,\theta,\phi,\psi)$ as 
\begin{eqnarray}\label{metric}
ds^2&=&-\left(dt-a\sin^2\theta d\phi-b\cos^2\theta d\psi\right)\nonumber\\&\times&\left[f\left(dt-a\sin^2\theta d\phi-b\cos^2\theta d\psi\right)\right.
\nonumber\\&+&\left.\frac{2q}{\Sigma}(b\sin^2\theta d\phi+a\cos^2\theta d\psi)\right]\nonumber\\&+&\Sigma\left(\frac{r^2dr^2}{\Delta}+d\theta^2\right)+\frac{\sin^2\theta}{\Sigma}\left[adt-(r^2+a^2)d\phi\right]^2\nonumber\\&+&\frac{\cos^2\theta}{\Sigma}\left[bdt-(r^2+b^2)d\psi\right]^2\nonumber\\&+&\frac{1}{r^2\Sigma}\left[a b dt-b(r^2+a^2)\sin^2\theta d\phi\right.\nonumber\\&-&\left. a(r^2+b^2)\cos^2\theta d\psi\right]^2\, ,
\end{eqnarray}
where we have set $\Lambda = 0$ and the metric coefficients are given by 
\begin{eqnarray}
f(r,\theta)&=&\frac{(r^2+a^2)(r^2+b^2)}{r^2\Sigma}-\frac{\mu\Sigma-q^2}{\Sigma^2},\nonumber\\
\Sigma(r,\theta)&=&r^2+a^2\cos^2\theta+b^2\sin^2\theta,\nonumber\\
\Delta(r)&=&(r^2+a^2)(r^2+b^2)+2abq+q^2-\mu r^2\ .
\end{eqnarray}
Here $a$ and $b$ are specific angular momenta parameters relative to two axes and they are related to angular momenta, $J_\phi, J_\psi$ as follows:  
\begin{eqnarray}
a+b=\frac{4}{\pi}\frac{J_{\phi}+J_{\psi}}{\mu+q}\, ,
\end{eqnarray}
with mass parameter $\mu=\frac{8M}{3\pi}$ and charge parameter $q=\frac{4Q}{\sqrt{3}\pi}$ of the black hole. The electromagnetic potential is given by
\begin{eqnarray}
\textbf{A}=\frac{-\sqrt{3}q}{2\Sigma}(dt-a\sin^2\theta d\phi-b\cos^2\theta d\psi)\, .
\end{eqnarray}

The horizon of the black hole follows from the relation $\Delta=0$, i.e. 
 \begin{eqnarray}\label{Eq:horizon}
 r_{\pm}&=&\pm\frac{\sqrt{\mu-2q-(a+b)^2}\pm\sqrt{\mu+2q-(a-b)^2}}{2}\, . \nonumber\\ 
 \end{eqnarray}
From the above expression it is evident that horizon does not exist unless the following inequalities: $ a^{2}+b^{2} +2|a||b|\leq \mu-2q$ and $ a^{2}+b^{2} - 2|a||b|\leq \mu+2q$ are satisfied. {Let's rewrite the horizon given in the above equation in terms of black hole mass, charge and angular momenta as 
 \begin{eqnarray}\label{Eq:horizon1}
 r_{+}&=&\frac{1}{4\sqrt{3\pi}\left(M+\frac{\sqrt{3} Q}{2}\right)}\left[\alpha \right.\nonumber\\&+&\left.\sqrt{\alpha^2+108 \pi J_{\phi} J_{\psi}+64 \sqrt{3}Q \left(M+\frac{\sqrt{3} Q}{2}\right)^2 }\right]\, ,\nonumber\\ 
 \end{eqnarray}
where 
\begin{eqnarray}
\alpha &=& \left(32M^3-27\pi\left(J_{\phi}+J_{\psi}\right)^2\right.\nonumber\\&-&\left. 72MQ^2-24\sqrt{3}~Q^3\right)^{1/2}\, . 
\end{eqnarray}
Note that black hole horizon exists if and only if $\alpha^2>0$, else it would be a naked singularity. Meanwhile, $\alpha=0$ corresponds to the extremal charged rotating black hole. }  
The area of the event horizon can be evaluated by setting $dr=dt=0$ and $r=r_{+}$ in the metric (\ref{metric}). The horizon metric reads as 
\begin{widetext}
\begin{eqnarray}
\label{det} g_{\alpha\beta}=
\left(\begin{array}{ccc} \Sigma & 0 & 0 \\ \\
0 & \left(r^2+a^2+\frac{a\left[a\left(\mu\Sigma-q^2\right)+2bq\Sigma\right]}{\Sigma}\sin^2\theta\right)\sin^2\theta & \frac{\left[a b\left(\mu\Sigma-q^2\right)+\left(a^2+b^2\right)q\Sigma\right]}{2\Sigma}\sin^2 2\theta\\ \\
0 & \frac{\left[a b\left(\mu\Sigma-q^2\right)+\left(a^2+b^2\right)q\Sigma\right]}{2\Sigma}\sin^2 2\theta & \left(r^2+b^2+\frac{b\left[b\left(\mu\Sigma-q^2\right)+2aq\Sigma\right]}{\Sigma}\cos^2\theta\right)\cos^2\theta \\ \\
 \end{array}\right)\, .
\end{eqnarray}
\end{widetext}
The horizon area is computed as 
\begin{eqnarray}
A=\int_{\Xi_{3}}\sqrt{det|g_{\alpha\beta}|}d\theta d\phi d\psi=
\frac{2\pi^2}{r_{+}}\left(\mu r_{+}^2-abq -q^2\right)\, ,
\end{eqnarray}
which must not decrease in any physical process according to the famous area non-decrease theorem~\cite{Grunau15}.

The angular velocity along $\phi$ and $\psi$ directions at the horizon $r=r_{+}$ are given by 
\begin{eqnarray}
\label{Eq:velocity1}
\Omega_{+}^{(\phi)}&=&\frac{a(r_+^2+b^2)+bq}{(r_+^2+a^2)(r_+^2+b^2)+abq}\, ,\\
\label{Eq:velocity2}
\Omega_{+}^{(\psi)}&=&\frac{b(r_+^2+a^2)+aq}{(r_+^2+a^2)(r_+^2+b^2)+abq}\, , 
\end{eqnarray}
for which the Killing field, $\chi=\chi^\alpha\partial_\alpha$ takes the form  
\begin{eqnarray}
\label{killingfield}\chi=\chi_{(t)}+\Omega_{+}^{(\phi)}\chi_{(\phi)}
+\Omega_{+}^{(\psi)}\chi_{(\psi)}\, .
\end{eqnarray}
Then surface gravity is defined by 
\begin{eqnarray}\label{sg1}
2k\chi_{\alpha}=\nabla_{\alpha}\left(-\chi_{\beta}\chi^{\beta}\right)|_{r=r_+}\,
,
\end{eqnarray}
or by
\begin{eqnarray}\label{sg2}
k^2=-\frac{1}{2}\left(\nabla_{\alpha}\chi_{\beta}\right)\left(\nabla^{\alpha}\chi^{\beta}\right)|_{r=r_+}\,
.
\end{eqnarray}
The surface gravity and electromagnetic potential at the horizon are respectively given by 
\begin{eqnarray}\label{surfacegravity}
k=\frac{\left(2r_{+}^2+a^2+b^2-\mu\right)r_{+}}{\mu r_{+}^2-abq-q^2}\, ,
\end{eqnarray}
and
\begin{eqnarray}
\Phi=-\chi^{\alpha}\textbf{A}_{\alpha}\vert_{r=r_{+}}=\frac{\sqrt{3}q r_{+}^2}{\mu r_{+}^2-abq-q^2}\, .
\end{eqnarray}

\section{Varitional identities and perturbation inequalities }\label{Sec:Perturbation}

It is well known that Lagrangian $L$ for a diffeomorphism covariant theory in $n$- dimensional manifold $\mathcal M$ can be described by metric $g_{\alpha\beta}$ with symmetrized covariant derivative and curvature tensor and other physical fields $\psi$~\cite{Wald94}. The variation of Lagrangian is then written as 
\begin{eqnarray}\label{Eq:Lag}
\delta L=E \delta\phi+d{\Theta}(\phi,\delta\phi),
\end{eqnarray}
where {we define all dynamical fields through $\phi=(g_{\alpha\beta},\psi)$ and $E$ as a parameter of Lagrangian, which consists of the fields $\phi$. Then equation of motion is given by ${E}=0$ while ${\Theta}$ represents symplectic potential $(n-1)$-form} and is written as 
\begin{eqnarray}
{\omega}(\phi,\delta_1\phi,\delta_2\phi)=\delta_1{\Theta}(\phi,\delta_2\phi)-\delta_2{\Theta}(\phi,\delta_1\phi)\, ,
\end{eqnarray}
where $\delta_{1,2}$ refers to the variations.   
The Noether current 5-form relative a vector field $\zeta^{\alpha}$ is defined by
\begin{eqnarray}\label{Eq:Noether}
{J}_\zeta={\Theta}(\phi,{L}_\zeta\phi)-\zeta \cdot{L},
\end{eqnarray}
for which $d{J}_\zeta=0$ is the equation of motion to be satisfied. According to \cite{Wald95}, one can define the Noether current in the following form 
\begin{eqnarray}\label{Eq:Noether1}
{ J}_\zeta=d{Q}_\zeta+ {C}_\zeta\, ,
\end{eqnarray}
where ${Q}_\zeta$ is referred to as the Noether charge while ${C}_\zeta=\zeta^{\alpha}{C}_{\alpha}$ is {the constraint of the theory--${C}_\zeta=0$ corresponds to the case when the equations of motion are satisfied}.

From the above equations (\ref{Eq:Noether}) and (\ref{Eq:Noether1}) for fixed $\zeta^{\alpha}$, we write the linear variational identity on a {Cauchy surface $\Xi$} 
\begin{eqnarray}\label{Eq:linear}
\int_{\partial\Xi}\delta {Q}_\zeta-\zeta\cdot{\Theta}(\phi,\delta\phi)&=&\int_{\Xi}{\omega}(\phi,\delta\phi,\mathcal{L}_\zeta\phi)\nonumber\\&-&\int_{\Xi}\zeta\cdot{E}\delta\phi-\int_{\Xi}\delta \mathbf{C}_\zeta\, ,
\end{eqnarray}
where the first term on the right is defined by %
\begin{eqnarray}\label{Eq:Ham}
\delta H_{\zeta}=\int_{\Xi}{\omega}(\phi,\delta\phi,\mathcal{L}_\zeta\phi)\, ,
\end{eqnarray}
which represents the variation of Hamiltonian associated with the vector field $\zeta^{\alpha}$. This reduces to $\delta H_{\zeta}=0$ if and only if $\zeta^{\alpha}$ is a Killing vector and a symmetry of $\phi$, thus satisfying both the equation of motion ${E}=0$ and $\mathcal{L}_\zeta\phi=0$. 
On the basis of linear variational identity, the non-linear one on the same surface is then defined by  
\begin{eqnarray}\label{Eq:non-linear}
\int_{\partial\Xi}\delta^2 {Q}_\zeta-\zeta\cdot\delta{\Theta}(\phi,\delta\phi)]&=&\int_{\Xi}{\omega}(\phi,\delta\phi,\mathcal{L}_\zeta\delta\phi)\nonumber\\ &-&\int_{\Xi}\zeta\cdot\delta{E}\delta\phi-\int_{\Xi}\delta^2 {C}_\zeta\, .\nonumber\\
\end{eqnarray}

Since $\zeta^{\alpha}$ is assumed to be a Killing field, Eq.~(\ref{Eq:linear}) for the linear variation reduces to
\begin{eqnarray}\label{Eq:E=0}
\int_{\partial\Xi}\delta {Q}_\chi-\chi\cdot{\Theta}(\phi,\delta\phi)&=&-\int_{\Xi}\delta \mathbf{C}_\chi\, , 
\end{eqnarray}
where $\chi^{\alpha}=\chi_{(t)}^{\alpha}+\Omega_{+}^{(\phi)}\chi_{(\phi)}^{\alpha}
+\Omega_{+}^{(\psi)}\chi_{(\psi)}^{\alpha}$ is the Killing vector with the horizon angular velocity $\Omega_{+}^{(\phi,\psi)}$. The Cauchy surface $\Xi$ defines the bifurcation surface $B$ at one end and spatial infinity at the other. Let us then rewrite the left-hand side of Eq.~(\ref{Eq:E=0}) on the Cauchy surface $\Xi$ 
\begin{eqnarray}\label{Eq:B+In}
\int_{\partial\Xi}\delta {Q}_\chi-\chi\cdot{\Theta}(\phi,\delta\phi)&=&\int_{\infty}\delta {Q}_\chi-\chi\cdot{\Theta}(\phi,\delta\phi)\nonumber\\&-&\int_{B}\delta {Q}_\chi-\chi\cdot{\Theta}(\phi,\delta\phi)\, . \nonumber\\
\end{eqnarray}
The contribution to boundary integral at infinity then yields 
\begin{eqnarray}\label{Eq:Boundary}
\int_{\infty}\delta {Q}_\chi-\chi\cdot{\Theta}(\phi,\delta\phi)&=&\delta M-\Omega_{+}^{(\phi)}\delta
J_{\phi}-\Omega_{+}^{(\psi)}\delta J_{\psi}\, \nonumber\\
\end{eqnarray}
with ADM mass $M$ and angular momenta $J_{\phi,\psi}$.
From Eqs.~(\ref{Eq:E=0}-\ref{Eq:Boundary}),  one can define the linear order variational identity (\ref{Eq:linear}) as
\begin{eqnarray}\label{Eq:linear1}
\delta M-\Omega_{+}^{(\phi)}\delta
J_{\phi}-\Omega_{+}^{(\psi)}\delta J_{\psi}&=&\int_B[\delta {Q}_\chi-\chi\cdot{\Theta}(\phi,\delta\phi)]\nonumber\\&-&\int_\Xi\delta {C}_\chi\, ,
\end{eqnarray}
for given {Cauchy surface $\Xi$} with a bifurcation surface $B$ on which the equation of motion is satisfied.   

On the other hand non-linear variational identity (\ref{Eq:non-linear}) then reads as 
\begin{eqnarray}\label{Eq:non-linear1}
\delta^2 M&-&\Omega_{+}^{(\phi)}\delta^2 J_{\phi}-\Omega_{+}^{(\psi)}\delta^2 J_{\psi}\nonumber\\&=&\int_B[\delta^2 {Q}_\chi-\chi\cdot\delta{\Theta}(\phi,\delta\phi)]\nonumber\\&-&\int_\Xi\chi\cdot\delta{E}\delta\phi-\int_\Xi\delta^2 {C}_\chi+\mathcal{E}_\Xi(\phi,\delta\phi)\, , 
\end{eqnarray}
where $\mathcal{E}_\Xi(\phi,\delta\phi)$ is canonical energy on the {Cauchy surface $\Xi$} as a non-linear correction to $\delta\phi$. 
 
For Eqs~(\ref{Eq:linear1}) and (\ref{Eq:non-linear1}), the symplectic potential 4-form is defined by  
\begin{eqnarray}\label{Eq:sym}
\Theta_{ijkh}\left(\phi,\delta\phi\right)&=&\frac{1}{16\pi}\epsilon_{ijkh\alpha}g^{\alpha\beta} g^{\gamma\eta}(\triangledown_{\eta}\delta g_{\beta\gamma}-\triangledown_{\beta}\delta g_{\gamma\eta})\nonumber\\&-&\frac{1}{4\pi}\epsilon_{ijkh\alpha}F^{\alpha\beta}\delta \textbf{A}_{\beta}\, , 
\end{eqnarray}
where the first term on the right is responsible for GR part while the second -- electromagnetic part as the Lagrangian is 
\begin{eqnarray}\label{Eq:Lag}
L=\frac{\epsilon}{16\pi}\left(R-F^{\alpha \beta}F_{\alpha\beta}\right)\, .
\end{eqnarray}
Hence we have 
\begin{eqnarray}
E(\phi)\delta\phi=-{\epsilon}(\frac{1}{2} T^{\alpha\beta}
\delta g_{\alpha \beta}+j^\alpha\delta \textbf{A}_{\alpha})\, ,
\end{eqnarray}
where $j^a=\frac{1}{4\pi} \triangledown_{b} F^{ab}$. Eq. (\ref{Eq:sym}) yields the corresponding symplectic current 
 \begin{eqnarray}
\omega_{ijkh}&=&\frac{1}{4\pi}\left[\delta_{2}(\epsilon_{ijkh\alpha } F^{\alpha\beta}) \delta_{1} \textbf{A}_{\beta}-\delta_{1}(\epsilon_{ijkh\alpha  } F^{\alpha\beta}) \delta_{2} \textbf{A}_{\beta}\right]\nonumber\\&+&\frac{1}{16\pi}\epsilon_{ijkh\alpha } w^{\alpha}\, ,
\end{eqnarray}
with
\begin{eqnarray}
w^{i}&=&P^{ijkh\alpha\beta}\left(\delta_{2} g_{jk} \triangledown_{h}\delta_{1}g_{\alpha\beta}-\delta_{1} g_{jk} \triangledown_{h}\delta_{2}g_{\alpha\beta}\right)\, ,\nonumber\\ 
P^{ijkh\alpha\beta}&=&g^{i\alpha} g^{\beta j} g^{k h}-\frac{1}{2}g^{ih} g^{j\alpha} g^{\beta k} - \frac{1}{2}g^{i j} g^{k h} g^{\alpha\beta}\nonumber\\ &-& \frac{1}{2}g^{j k} g^{i\alpha} g^{\beta h} + \frac{1}{2}g^{jk} g^{i h} g^{\alpha\beta}\, .
 	\end{eqnarray}
Taking into account $\mathcal{L}_{\zeta} g_{\alpha\beta} = \triangledown_{\alpha}\zeta_{\beta} +\triangledown_{\beta}\zeta_{\alpha}$ and $\triangledown_{\alpha}\textbf{A}_{\beta}=F_{\alpha\beta}+\triangledown_{\beta}\textbf{A}_{\alpha}$, the Noether current 4-form is given by 
\begin{eqnarray}\label{Eq:Noether2}
(J_{\zeta})_{ijkh}&=&\frac{1}{8\pi}\epsilon_{ijkh\alpha } \triangledown_{\beta}(\triangledown^{[\beta} \chi^{\alpha]}) + \epsilon_{ijkh\alpha } T_{\beta}^{\alpha} \zeta^{\beta}\nonumber\\&+&\frac{1}{4\pi}\epsilon_{ijkh \alpha }\triangledown_{\gamma}(F^{\gamma\alpha} \textbf{A}_{\beta} \zeta^{\beta}) + \epsilon_{ijkh\alpha } \textbf{A}_{\beta} j^{\alpha} \chi^{\beta}\, , \nonumber\\ 
\end{eqnarray}
as well as the Noether charge $Q_{\zeta}$ and the constraint $C_{\zeta}$ read as 
\begin{eqnarray}
(Q_{\zeta})_{ijk}&=&-\frac{1}{16\pi}\epsilon_{ijk \alpha\beta}\triangledown^{\alpha}\zeta^{\beta}-\frac{1}{8\pi}\epsilon_{ijk \alpha\beta} F^{\alpha\beta } \textbf{A}_{\gamma} \zeta^{\gamma}\, \nonumber\\
(C_{\gamma})_{ijkh}&=&\epsilon_{ijkh \alpha}(T_{\gamma}^{\alpha} + \textbf{A}_{\gamma} j^{\alpha})\, . 
\end{eqnarray}

\section{Over extremalizing black hole via gedanken experiments }\label{Sec:extremal}

\subsection{Extremal case}

Here we consider a particle absorption by an extremal black hole of mass $M$, angular momenta $J_{\psi}$ and $J_{\phi}$ and electric charge $Q$.  {From Eq. (7), the extremality condition reads as
\begin{eqnarray}\label{Eq:extremal-state}
32M^3=27\pi\left(J_{\phi}+J_{\psi}\right)^2+72MQ^2+24\sqrt{3}~Q^3\, .
\end{eqnarray}
A particle of energy $\delta M$ and angular momenta $\delta J_{\psi}$ and $\delta J_{\phi}$ and charge $\delta Q$ is thrown into black hole horizon. This leads to increase in the corresponding parameters of black hole, and a perturbed stationary state would be attained with parameters, $M + \delta M$, $J + \delta J_{\phi}$, $J + \delta J_{\psi}$, and $Q+\delta Q$.
The condition for over-extremalization or WCCC violation would require the following inequality  
\begin{eqnarray}
96M^2\delta M&<&54\pi\left(J_{\phi}+J_{\psi}\right)\left(\delta J_{\phi}+\delta J_{\psi}\right)+72Q^2\delta M\nonumber\\&+&144MQ\delta Q +72\sqrt{3}~Q^2\delta Q\, ,
\end{eqnarray}
for the first order linear accretion. An extremal black hole will be pushed to over-extremal state if and only if 
\begin{eqnarray}\label{over-extremal}
\delta M&-&\frac{9\pi\left(J_{\phi}+J_{\psi}\right)}{4(4M^2-3Q^2)}\left(\delta J_{\phi}+\delta J_{\psi}\right)\nonumber\\&-&\frac{3\left(2MQ-\sqrt{3}Q^2\right)}{\left(4M^2-3Q^2\right)}\delta Q<0\, .
\end{eqnarray}
}

{We should then examine whether over-extremal state satisfying Eq.~(\ref{over-extremal}) occurs or not?  Let's suppose that a black hole with initial given state is bombarded by test particles of appropriate parameters described by the stress-energy tensor $T_{\alpha\beta}$. Consequently, black hole parameters are increased by following amounts~\cite{Sorce-Wald17} 
\begin{eqnarray}
\label{change_Mass1} \delta M&=&
\int_{H}\epsilon_{ijkh \alpha}\chi_{(t)}^{\gamma}\left(\delta T_{\gamma}^{\alpha} + \textbf{A}_{\gamma} \delta j^{\alpha}\right)\, ,\\
\label{change_angular1} \delta J_{\phi}&=&
-\int_{H}\epsilon_{ijkh \alpha}\chi_{(\phi)}^{\gamma}\left(\delta T_{\gamma}^{\alpha} + \textbf{A}_{\gamma} \delta j^{\alpha}\right) \, , \\
\label{change_angular2}\delta J_{\psi}&=&
-\int_{H}\epsilon_{ijkh \alpha}\chi_{(\psi)}^{\gamma}\left(\delta T_{\gamma}^{\alpha} + \textbf{A}_{\gamma} \delta j^{\alpha}\right)\, ,
\end{eqnarray}
where the integration is over surface element on the event horizon $r_{+}$. We assume that at the end of the process, black
hole attains another stationary state. {Since the term $\int_B[\delta {Q}_\chi-\chi\cdot{\Theta}(\phi,\delta\phi)]$ vanishes because of no perturbation at the bifurcation surface \cite{Sorce-Wald17},  Eq.~(\ref{Eq:linear1}) then yields }
\begin{eqnarray}
\label{Eq:change_Mass2} \delta M&-&\Omega_{+}^{(\phi)}\delta
J_{\phi}-\Omega_{+}^{(\psi)}\delta J_{\psi}=-\int_\Xi\delta {C}_\gamma=
\nonumber\\
&-&\int_{H}\epsilon_{ijkh \alpha}\left(\chi_{(t)}^{\gamma}+\Omega_{+}^{(\phi)}\chi_{(\phi)}^{\gamma}+\Omega_{+}^{(\psi)}\chi_{(\psi)}^{\gamma}\right)\nonumber\\
&\times &\left(\delta T_{\gamma}^{\alpha} + \textbf{A}_{\gamma} \delta j^{\alpha}\right)\, ,
\end{eqnarray}
where $\chi^{\gamma}$ is null generator of the horizon $r_+$. {Eq.~(\ref{Eq:change_Mass2}) ensures that particle
crossed the horizon eventually.} Bearing in mind $\Phi=-\chi^{\gamma}\textbf{A}_{\gamma}\vert_{r=r_{+}}$ and using $\int_{H}\delta(\epsilon_{ijkh \alpha}j^{\alpha})=\delta
Q$ for the perturbed charge fallen into the horizon $r_{+}$, we rewrite  Eq.~(\ref{Eq:change_Mass2}) as
\begin{eqnarray}
\label{Eq:change_Mass3} \delta M&-&\Omega_{+}^{(\phi)}\delta
J_{\phi}-\Omega_{+}^{(\psi)}\delta J_{\psi}-\Phi_{+}\delta
Q\nonumber\\& =& - \int_{H}\epsilon_{ijkh \alpha} \chi_{\gamma} \delta T^{\gamma\alpha}\, ,
\end{eqnarray}
where volume element on the horizon is written as $\epsilon_{ijkh \alpha}=-5\tilde{\epsilon}_{[ijkh} k_{\alpha ]}$ We then write 
\begin{eqnarray}
- \int_{H}\epsilon_{ijkh \alpha} \chi_{\gamma} \delta T^{\gamma\alpha}=\int_{H}\tilde{\epsilon}_{ijkh} \chi_{\gamma}k_{\alpha} \delta T^{\gamma\alpha}\, .
\end{eqnarray}
This clearly shows that the right hand side is only positive only when the null energy condition is satisfied, i.e. $\delta T_{\alpha\beta}k^{\alpha}k^{\beta}\geq 0$. This leads to the inequality  
\begin{eqnarray}\label{Eq:change_Mass4} 
\delta M&-&\Omega_{+}\left(\delta
J_{\phi}+\delta J_{\psi}\right)-\Phi_{+}\delta
Q\geq 0\, .
\end{eqnarray}

For the extremal black hole we have 
\begin{eqnarray} 
\Omega_{+}&=&\frac{9\pi\left(J_{\phi}+J_{\psi}\right)}{4(4M^2-3Q^2)}\, , \\
\Phi_{+}&=&\frac{3\left(2MQ-\sqrt{3}Q^2\right)}{\left(4M^2-3Q^2\right)}\, 
\end{eqnarray}
\\
and the inequality (\ref{Eq:change_Mass4}) becomes 
\begin{eqnarray}\label{Eq:change_Mass5} 
\delta M&-&\frac{9\pi\left(J_{\phi}+J_{\psi}\right)}{4(4M^2-3Q^2)}\left(\delta
J_{\phi}+\delta J_{\psi}\right)\nonumber\\&-&\frac{3\left(2MQ-\sqrt{3}Q^2\right)}{\left(4M^2-3Q^2\right)}\delta
Q\geq 0\, .
\end{eqnarray}
This inequality clearly contradicts the inequality (\ref{over-extremal}). Thus an extremal black hole cannot be overspun and WCCC holds. 

Further we must show that new perturbed state is also indeed extremal. To ensure that it is indeed not possible to {over-extremalize an
extremal black hole}. From the first law of black hole dynamics we write }
\begin{eqnarray}
\label{firstlaw}\delta M= \frac{k}{8\pi}\delta
A+\Omega^{(\phi)}\delta J_{\phi}+\Omega^{(\psi)}\delta
J_{\psi}+\Phi\delta
Q\, ,
\end{eqnarray}
where $M=M(A,J_{\phi},J_{\psi},Q)$ and horizon area, $A=A(J_{\phi},J_{\psi},Q)$. For extremal black hole, we will consider variation in the
mass
\begin{eqnarray}
\label{Eq:mass1}\delta M_{ext}&=& \left(\frac{\partial
M}{\partial A}\frac{\partial A_{ext}}{\partial
J_{\phi}}+\frac{\partial M}{\partial
J_{\phi}}\right)\delta J_{\phi}\nonumber\\
&+&\left(\frac{\partial M}{\partial A}\frac{\partial
A_{ext}}{\partial J_{\psi}}+\frac{\partial M}{\partial
J_{\psi}}\right)\delta
J_{\psi}\nonumber\\
&+&\left(\frac{\partial
M}{\partial A}\frac{\partial A_{ext}}{\partial
Q}+\frac{\partial M}{\partial
Q}\right)\delta Q\nonumber\\
&=&\frac{k}{8\pi}\delta A + \Omega_{+}^{(\phi)}\delta J_{\phi}+\Omega_{+}^{(\psi)}\delta
J_{\psi}+\Phi_{+}\delta
Q\, ,
\end{eqnarray}
{where 
\begin{eqnarray}\label{Eq:sur-gravity}
k&=&\frac{\partial
M}{\partial A}\, ,\\
\delta A&=&\frac{\partial A_{ext}}{\partial J_{\phi}}\delta
J_{\phi}+\frac{\partial A_{ext}}{\partial J_{\psi}}\delta
J_{\psi}+\frac{\partial A_{ext}}{\partial Q}\delta
Q\, .\label{Eq:hor-area}
\end{eqnarray}}
The surface gravity goes to zero $k\rightarrow0$ for an extremal black hole. As a result, Eq.~(\ref{Eq:mass1}) yields
\begin{eqnarray}
\label{Eq:mass2}\delta M_{ext}=\Omega_{+}\left(\delta
J_{\phi}+\delta J_{\psi}\right)+\Phi_{+}\delta Q\, ,
\end{eqnarray}
 {which characterizes an extremal black hole
$M=M_{ext}(J_{\phi},J_{\psi},Q)$. The black hole exists provided $M \geq M_{ext}(J_{\phi},J_{\psi},Q)$, and if opposite is the case, $M < M_{ext}(J_{\phi},J_{\psi},Q)$, over-extremal state occurs.  If a particle with angular momenta and charge crosses the
horizon of an extremal black hole which results into black hole's angular momenta and charge enhanced to $J_{\phi}+\delta J_{\phi}$,
$J_{\psi}+\delta J_{\psi}$ and $Q+\delta Q$. } In view of Eqs~(\ref{Eq:change_Mass4}) and (\ref{Eq:mass2}), we then write final mass  is given
\begin{eqnarray}
\label{Eq:finalmass}M+\delta M&\geq &
M+\Omega_{+}\left(\delta
J_{\phi}+\delta J_{\psi}\right)+\Phi_{+}\delta Q\nonumber\\&=& M_{ext}(J_{\phi},J_{\psi},Q)+\delta
M_{ext}\nonumber\\
&=&M_{ext}(J_{\phi}+\delta J_{\phi},J_{\psi}+\delta J_{\psi},Q+\delta Q)\, .
\end{eqnarray}
As is clear from the above equation that final black hole mass is not less than the initial extremal mass and hence it has not been over extremalized. All this is in agreement with the third law of black hole thermodynamics \cite{Bardeen73b,Wald74b,Israel86,Dadhich97}. Thus an extremal black hole cannot be converted into an over extremal state, and there occurs no violation of WCCC. 

Next, we investigate over-extremal state for a near-extremal black hole for linear and non-linear perturbations through gedanken experiments.

\subsection{Near-extremal case }\label{Sec:gedanken}

In this subsection we apply new gedanken experiment developed by the Sorce and Wald~\cite{Sorce-Wald17} to over-extremalize near extremal black hole. According to the gedanken experiment one should take into account a one-parameter family of field $\phi(\lambda)$ and the background spacetime is characterized by $T_{\alpha\beta}=0$ and $j^{\alpha}=0$. For this we have already considered a hypersurface as $\Xi=\Xi_{1}\cup H$ endowed with specific properties. So this hypersurface contains such a region from which bifurcation surface $B$ starts and continues up the horizon portion $H$ of $\Xi$ till it becomes spacelike $\Xi_{1}$. After that it reaches spatial infinity to become asymptotically flat. Based on the particular characteristics of the $\Xi$, we work on the second order variational identity for a near extremal black hole. Let us recall  Eq.~(\ref{Eq:non-linear1})
\begin{eqnarray}\label{Eq:Non1}
\delta^2 M&-&\Omega_{+}^{(\phi)}\delta^2 J_{\phi}-\Omega_{+}^{(\psi)}\delta^2 J_{\psi}=\int_B[\delta^2 {Q}_\chi-\chi\cdot\delta{\Theta}(\phi,\delta\phi)]\nonumber\\&-&\int_\Xi\chi\cdot\delta{E}\delta\phi-\int_\Xi\delta^2 {C}_\chi+\mathcal{E}_\Xi(\phi,\delta\phi)\nonumber\\ 
&=& \int_B[\delta^2 {Q}_\chi-\chi\cdot\delta{\Theta}(\phi,\delta\phi)]+\mathcal{E}_H(\phi,\delta\phi)\nonumber\\&-&\int_H\chi\cdot\delta{E}\delta\phi\nonumber\\&-&\int_{H}\epsilon_{ijkh \alpha}\left(\chi_{(t)}^{\gamma}+\Omega_{+}^{(\phi)}\chi_{(\phi)}^{\gamma}+\Omega_{+}^{(\psi)}\chi_{(\psi)}^{\gamma}\right)\nonumber\\
&\times &\left(\delta^2 T_{\gamma}^{\alpha} + \textbf{A}_{\gamma} \delta^2 j^{\alpha}\right)\nonumber\\ &=& \int_B[\delta^2 {Q}_\chi-\chi\cdot\delta{\Theta}(\phi,\delta\phi)]+\mathcal{E}_H(\phi,\delta\phi)\nonumber\\&+& \int_{H}\tilde{\epsilon}_{ijkh} \chi_{\gamma}k_{\alpha} \delta^2 T^{\gamma\alpha}+\Phi_{+}\delta^2
Q\, , 
\end{eqnarray}
where $\chi^{\alpha}$ is tangent to $H$ and applied the gauge condition $\chi^{\alpha}\delta \textbf{A}_{\alpha}=0$ on $H$. In the last step, we impose the null energy condition $\delta^2 T_{\alpha\beta}k^{\alpha}k^{\beta}\geq0$ to rewrite the above equation 
\begin{eqnarray}\label{Non2}
\delta^2 M&-&\Omega_{+}^{(\phi)}\delta^2 J_{\phi}-\Omega_{+}^{(\psi)}\delta^2 J_{\psi}-\Phi_{+}\delta^2
Q\nonumber\\ &=& \int_B[\delta^2 {Q}_\chi-\chi\cdot\delta{\Theta}(\phi,\delta\phi)]+\mathcal{E}_H(\phi,\delta\phi)\, . 
\end{eqnarray}  

Let us then evaluate the first and second terms on the right-hand side of Eq.~(\ref{Non2}) and rewrite these terms for a one-parameter fieled $\phi^{MGS}(\lambda)$, 
\begin{eqnarray}
\int_B[\delta^2 {Q}_\chi-\chi\cdot\delta{\Theta}(\phi,\delta\phi^{MGS})] ~~\mbox{and}~~\mathcal{E}_{H}(\phi,\delta\phi^{MGS})\, ,
\end{eqnarray}
where $\delta\phi^{MGS}$ is the perturbation caused by falling in matter to the minimally gauged supergravity black hole with following parameters 
\begin{eqnarray}\label{Eq:MJQ}
M(\lambda)&=& M+\lambda\delta M\, ,\nonumber\\
J_{\phi}(\lambda)&=&J_{\phi}+\lambda\delta J_{\phi}\, , \nonumber\\
J_{\psi}(\lambda)&=&J_{\psi}+\lambda\delta J_{\psi}\, , \nonumber\\
Q(\lambda)&=&Q+\lambda\delta Q\, . 
\end{eqnarray}
Note here that we choose $\delta M$, $\delta Q$, and $\delta J_{\phi,\psi}$ in such a way that they are consistent with the linear order perturbation Eq.~(\ref{Eq:change_Mass4}). However, $\delta^2 M=\delta^2 J_{\phi,\psi}=\delta^2 Q_{B}=\delta E=\mathcal{E}_{H}(\phi,\delta\phi^{MGS})=0$ is satisfied for this one parameter family of fields. Thus, by imposing the condition $\chi^{\alpha}=0$ at the bifurcation surface $B$ we have      
\begin{eqnarray}
\delta^2 M&-&\Omega_{+}^{(\phi)}\delta^2 J_{\phi}-\Omega_{+}^{(\psi)}\delta^2 J_{\psi}-\Phi_{+}\delta^2
Q\nonumber\\&=&\int_B[\delta^2 {Q}_\chi-\chi\cdot\delta{\Theta}(\phi,\delta\phi^{MGS})]\nonumber\\&\geq & -\frac{k}{8\pi}\delta^2 A^{MGS}\, .
\end{eqnarray}
This is the non-linear variational identity for the one-parameter family of perturbation.  

{Following all the above procedure we apply this new version of gedanken experiment to probe over-extremalization of near extremal  black hole. Let us recall the extremality condition Eq~(\ref{Eq:extremal-state}),}
$$
32M^3-27\pi\left(J_{\phi}+J_{\psi}\right)^2-72MQ^2-24\sqrt{3}~Q^3=0\, .
$$
Thus a near extremal state is characterized as  
\begin{eqnarray} \label{Eq:function}
f(\lambda)&=& 32M(\lambda)^3-27\pi\left[J_{\phi}(\lambda)+J_{\psi}(\lambda)\right]^2\nonumber\\&-&72M(\lambda)Q(\lambda)^2-24\sqrt{3}~Q(\lambda)^3\, ,\end{eqnarray}
where $f(0)=\alpha^2$, being a bit larger than zero, and {$M(\lambda)$, $J_{\phi}(\lambda)$, $J_{\psi}(\lambda)$ and $ Q(\lambda)$ are as defined by Eq.~(\ref{Eq:MJQ})}.  
{To jump from sub-extremal to over-extremal state we must obtain $f(\lambda)<0$}, and for that we now expand $f(\lambda)$ up to second order in $\alpha$ and $\lambda$ as                                       
 %
 \begin{eqnarray}
f(\lambda)=\alpha^2+f_1\lambda+f_2\lambda^2+O(\lambda^3, \lambda^2\alpha, \lambda\alpha^2,\alpha^3),
\end{eqnarray}
where
 \begin{widetext}
 \begin{eqnarray}\label{Eq:liner}
f_1&=& 24\left(4M^2-3Q^2\right)\left[\delta M- \frac{9\pi\left(J_{\phi}+J_{\psi}\right)}{4(4M^2-3Q^2)}\left(\delta J_{\phi}+\delta J_{\psi}\right)-\frac{3\left(2MQ-\sqrt{3}Q^2\right)}{\left(4M^2-3Q^2\right)}~\delta Q \right] \, ,\\
f_2&=&\left\{12\left(4M^2-3Q^2\right)\left[\delta^2 M-\frac{9\pi\left(J_{\phi}+J_{\psi}\right)}{4(4M^2-3Q^2)}\left(\delta^2 J_{\phi}+\delta^2 J_{\psi}\right)-\frac{3\left(2MQ-\sqrt{3}Q^2\right)}{\left(4M^2-3Q^2\right)}~\delta^2 Q\right]\right.\nonumber\\&+&\left. 96M(\delta M)^2-27\pi\left(\delta J_{\phi}+\delta J_{\psi}\right)^2+72\left(M(\delta Q)^2+2Q\delta M\delta Q +\sqrt{3}Q(\delta Q)^2\right)\right\}\, . \label{Eq:nonlinear}    
\end{eqnarray}
\end{widetext}
In Eq.~(\ref{Eq:liner}), the expression in the bracket is written for optimal choice of linear order correction
\begin{widetext}
\begin{eqnarray}\label{Eq:optimal}
\delta M &-&\frac{9\pi\left(J_{\phi}+J_{\psi}\right)}{4(4M^2-3Q^2)}\left(\delta J_{\phi}+\delta J_{\psi}\right)+\frac{3\left(2MQ-\sqrt{3}Q^2\right)}{\left(4M^2-3Q^2\right)}~\delta Q =\nonumber\\&-&\frac{\sqrt{27 \pi  J_{\phi} J_{\psi}+4 \sqrt{3} Q \left(2 M+\sqrt{3} Q\right)^2}}{\left(27 \pi  J_{\phi} J_{\psi}+4 \sqrt{3} Q \left(2 M+\sqrt{3} Q\right)^2\right)^2 \left(9\pi(J_{\phi}+J_{\psi})^2+\frac{4\sqrt{3}}{3}Q\left(2 M+\sqrt{3} Q\right)^2\right)^2}\nonumber\\ &\times & \left[ 6 \pi  \left(M+\frac{\sqrt{3} Q}{2}\right) \left(144 \sqrt{3} \pi  Q \left(M+\frac{\sqrt{3} Q}{2}\right)^2 \left[\delta J_{\psi} J_{\phi}^3+2 J_{\psi} J_{\phi}^2 (\delta J_{\phi}+2 \delta J_{\psi})+2 J_{\phi} J_{\psi}^2 (\delta J_{\psi}+2 \delta J_{\phi})+\delta J_{\phi} J_{\psi}^3\right]\right.\right.\nonumber\\&+& \left. 243 \pi ^2 {J_{\phi}} {J_{\psi}} ({J_{\phi}}+{J_{\psi}})^2 (\delta J_{\psi} {J_{\phi}}+\delta J_{\phi} {J_{\psi}})+16 Q^2 \left(2 M+\sqrt{3} Q\right)^4 \big[{J_{\phi}} (\delta J_{\phi}+2 \delta J_{\psi})+{J_{\psi}} (\delta J_{\psi}+2 \delta J_{\phi})\big]\right)\nonumber\\&+& \left. 256 Q^2 \left(M+\frac{\sqrt{3} Q}{2}\right)^4 \left(9 \sqrt{3} \pi  {J_{\phi}} {J_{\psi}}+4 Q \left(2 M+\sqrt{3} Q\right)^2\right)\delta Q \right]\mathbf{\alpha}\, .
\end{eqnarray}
\end{widetext}

\subsection{With two rotations}


\subsubsection {Linear order accretion}

In view of the above equation ~(\ref{Eq:optimal}), we rewrite $f(\lambda)$ for linear order correction as
\begin{widetext}
\begin{eqnarray}\label{Eq:linear-order}
f(\lambda)&=&\alpha^2-\frac{6\left(2 M+\sqrt{3} Q\right)^{-1}}{\left(27 \pi  J_{\phi} J_{\psi}+4 \sqrt{3} Q \left(2 M+\sqrt{3} Q\right)^2\right)^{1/2} \left(9\pi(J_{\phi}+J_{\psi})^2+\frac{4\sqrt{3}}{3}Q\left(2 M+\sqrt{3} Q\right)^2\right)^2}\nonumber\\ &\times & \left[ 6 \pi  \left(M+\frac{\sqrt{3} Q}{2}\right) \left(144 \sqrt{3} \pi  Q \left(M+\frac{\sqrt{3} Q}{2}\right)^2 \left[\delta J_{\psi} J_{\phi}^3+2 J_{\psi} J_{\phi}^2 (\delta J_{\phi}+2 \delta J_{\psi})+2 J_{\phi} J_{\psi}^2 (\delta J_{\psi}+2 \delta J_{\phi})+\delta J_{\phi} J_{\psi}^3\right]\right.\right.\nonumber\\&+& \left. 243 \pi ^2 {J_{\phi}} {J_{\psi}} ({J_{\phi}}+{J_{\psi}})^2 (\delta J_{\psi} {J_{\phi}}+\delta J_{\phi} {J_{\psi}})+16 Q^2 \left(2 M+\sqrt{3} Q\right)^4 \big[{J_{\phi}} (\delta J_{\phi}+2 \delta J_{\psi})+{J_{\psi}} (\delta J_{\psi}+2 \delta J_{\phi})\big]\right)\nonumber\\&+& \left. 256 Q^2 \left(M+\frac{\sqrt{3} Q}{2}\right)^4 \left(9 \sqrt{3} \pi  {J_{\phi}} {J_{\psi}}+4 Q \left(2 M+\sqrt{3} Q\right)^2\right)\delta Q \right]~\bf{\alpha}~\bf{\lambda} +\mathcal O(\lambda^2)\, ,
\end{eqnarray}
\end{widetext}
from which it is evident that it is always possible to obtain $f(\lambda)<0$ for suitable values of given parameters. Thus black hole could be over-extremalized. To ensure this, we try to explore $f(\lambda)$ numerically. {From Eq.~(\ref{Eq:horizon}), the extremal condition $\mu-2q=(a+b)^2$ yields  
\begin{eqnarray}\label{Eq:Extremal}
\sqrt{\frac{32}{27\pi}\left(M-\sqrt{3}Q\right)}=\frac{J_{\phi}+J_{\psi}}{M+\frac{\sqrt{3}}{2}Q}\, .
\end{eqnarray}
From Eq.~(\ref{Eq:Extremal}) it is clear that a near-extremality requires $Q^2 < M^2/3 $, which in turn allows us to choose $Q=0.5M$. For given $Q=0.5 M$, $f(0)=\alpha^2$ corresponding to the near extremality defines the angular momenta numerically, $J_{\phi}+J_{\psi}=0.322011$ for the given value $\alpha=0.01$. For this thought experiment one can take different values of black hole parameters and even smaller values of $\alpha$. Setting $M=1$, let's choose $\delta J_{\phi}=0.001\ll J_{\phi} $, $\delta J_{\psi}=0.001\ll J_{\psi} $ and $\delta Q=0.003 \ll Q$ in order for the test particle approximation to remain valid. Let's now evaluate Eq.~(\ref{Eq:linear-order}) numerically, thereby $f(0.1)=-0.00045<0$. That is, it could be over-extremalized under linear order accretion. It thus indicates violation of WCCC at the linear order. The obtained numerical results are shown in Fig.~\ref{fig1}. 
\begin{figure}
\centering
  \includegraphics[width=0.45\textwidth]{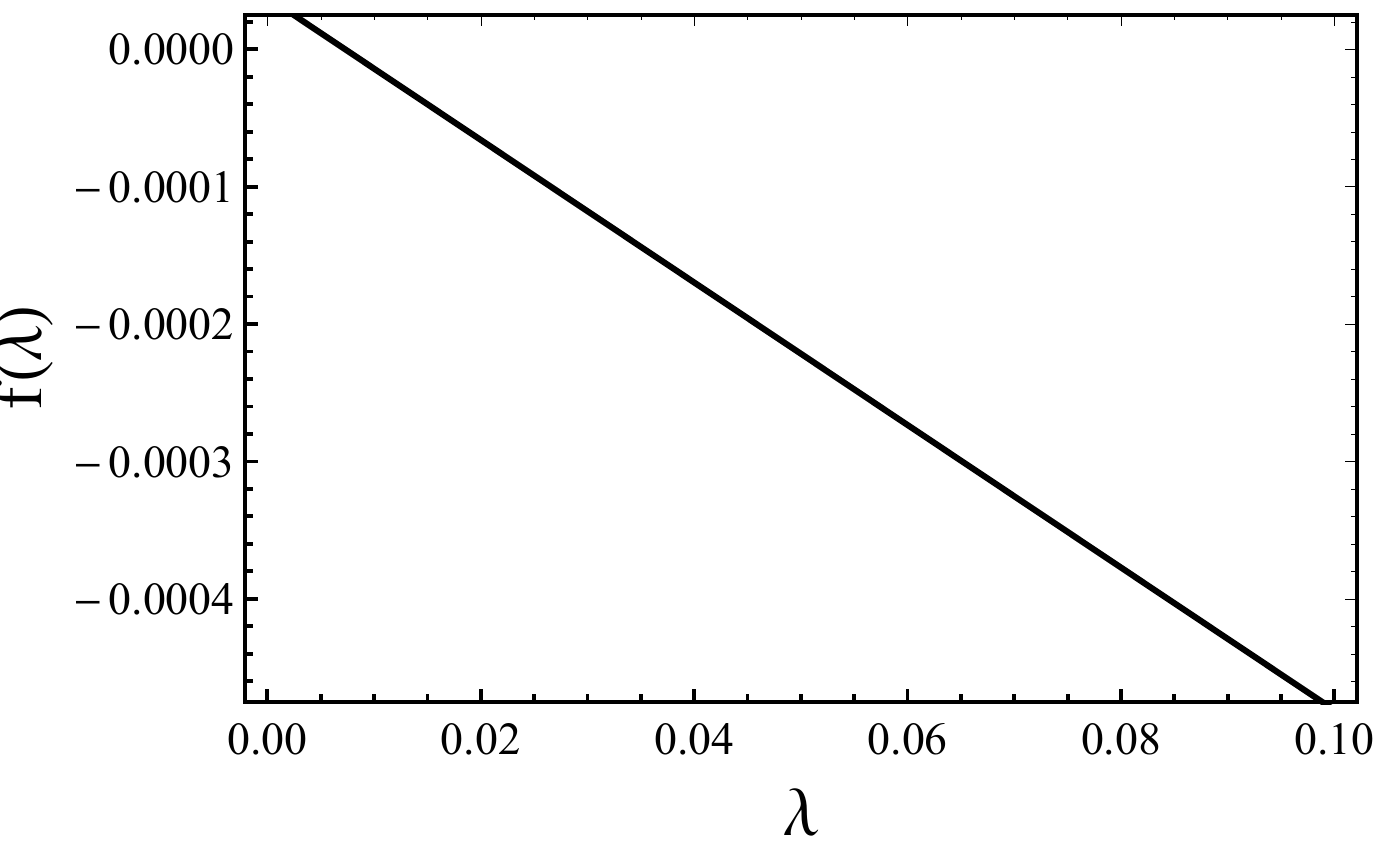}

\caption{\label{fig1} $f(\lambda)$ against $\lambda$ for the given values of test particle and black hole parameters. }
\end{figure}

\subsubsection {Non-linear order accretion}                

We here consider the second order particle accretion $O(\lambda^2)$ so as to understand what might happen in the case of non-linear regime. Let's start from  Eq.~(\ref{Eq:nonlinear}), where the non-linear terms are given by   
\begin{widetext}
\begin{eqnarray}\label{Eq:second-order}
&&\delta^2 M-\frac{9\pi\left(J_{\phi}+J_{\psi}\right)}{4(4M^2-3Q^2)}\left(\delta^2 J_{\phi}+\delta^2 J_{\psi}\right)-\frac{3\left(2MQ-\sqrt{3}Q^2\right)}{\left(4M^2-3Q^2\right)}~\delta^2 Q \geq -\frac{k}{8\pi}\delta^2 A= \frac{1}{12\left(4M^2-3Q^2\right)\alpha^2}\nonumber\\
&&\times\Bigg(N_{1}\left(M,Q,J_{\phi},J_{\psi}\right)\delta M^2+N_{2}\left(M,Q,J_{\phi},J_{\psi},\delta J_{\phi},\delta J_{\psi}\right)\delta M+N_{3}\left(M,Q,J_{\phi}\right)\delta J_{\psi}^2+N_{4}\left(M,Q,J_{\phi},J_{\psi}\right)\delta J_{\phi}\delta J_{\psi}\nonumber\\&&+N_{5}\left(M,Q,J_{\phi},J_{\psi},\delta J_{\phi},\delta J_{\psi}\right)\delta M\delta Q+N_{6}\left(M,Q,J_{\psi}\right)\delta J_{\phi}^2+N_{7}\left(M,Q,J_{\phi},J_{\psi},\delta J_{\phi},\delta J_{\psi}\right)\delta Q+N_{8}\left(M,Q,J_{\phi},J_{\psi}\right)\delta Q^2\Bigg)\, . \nonumber\\
\end{eqnarray}
\end{widetext}
Here the function $N_{i}$ is related to the black hole parameters in a complicated way. When we take into account non-linear term $O(\lambda^2)$ by using Eq.~(\ref{Eq:second-order}) and optimal choice of linear order correction, the function $f(\lambda)$ takes the form 
\begin{widetext}
\begin{eqnarray}\label{Eq:both}
f(\lambda)&> &\left(\alpha-\frac{3\left(2 M+\sqrt{3} Q\right)^{-1}~\lambda }{\left(27 \pi  J_{\phi} J_{\psi}+4 \sqrt{3} Q \left(2 M+\sqrt{3} Q\right)^2\right)^{1/2} \left(9\pi(J_{\phi}+J_{\psi})^2+\frac{4\sqrt{3}}{3}Q\left(2 M+\sqrt{3} Q\right)^2\right)^2}\right.\nonumber\\ &\times & \left[ 6 \pi  \left(M+\frac{\sqrt{3} Q}{2}\right) \left(144 \sqrt{3} \pi  Q \left(M+\frac{\sqrt{3} Q}{2}\right)^2 \left[\delta J_{\psi} J_{\phi}^3+2 J_{\psi} J_{\phi}^2 (\delta J_{\phi}+2 \delta J_{\psi})+2 J_{\phi} J_{\psi}^2 (\delta J_{\psi}+2 \delta J_{\phi})+\delta J_{\phi} J_{\psi}^3\right]\right.\right.\nonumber\\&+& \left. 243 \pi ^2 {J_{\phi}} {J_{\psi}} ({J_{\phi}}+{J_{\psi}})^2 (\delta J_{\psi} {J_{\phi}}+\delta J_{\phi} {J_{\psi}})+16 Q^2 \left(2 M+\sqrt{3} Q\right)^4 \big[{J_{\phi}} (\delta J_{\phi}+2 \delta J_{\psi})+{J_{\psi}} (\delta J_{\psi}+2 \delta J_{\phi})\big]\right)\nonumber\\&+& \left. \left.256 Q^2 \left(M+\frac{\sqrt{3} Q}{2}\right)^4 \left(9 \sqrt{3} \pi  {J_{\phi}} {J_{\psi}}+4 Q \left(2 M+\sqrt{3} Q\right)^2\right)\delta Q \right]\right)^2+\mathcal O(\alpha^3,\alpha^2\lambda,\alpha\lambda^2,\lambda^{3})\, .
\end{eqnarray}
\end{widetext}

This clearly shows $f(\lambda) > 0$ always.   Thus, it verifies the expected result that a five dimensional charged rotating black hole in minimally gauged supergravity cannot be over extremalized for a non-linear order accretion while the opposite is true for a linear order accretion. Under non-linear accretion WCCC is therefore always obeyed.
 
%

\subsection{With single rotation}

\subsubsection{Linear order accretion}

Let's consider a particular case of single rotation, for which Eq.~(\ref{Eq:linear-order}) takes the following form 
%
\begin{eqnarray} \label{Eq:linear-order1}
f(\lambda)&=&\alpha^2-\frac{48\times 3^{3/4}Q^{3/2}\left(2 M+\sqrt{3} Q\right)^3}{\left(9\sqrt{3}\pi J_{\psi}^2+4Q\left(2 M+\sqrt{3} Q\right)^2\right)^2}\nonumber\\
&\times &\bigg(3\pi J_{\psi}\delta J_{\psi}+4Q\left(2M+\sqrt{3}Q\right)\delta Q\bigg)\alpha \lambda +\mathcal O(\lambda^2)\, .\nonumber\\
\end{eqnarray}
%

It is clear from the above equation that overspinning/charging is quite possible in general. However let's consider various cases separately.
\begin{itemize}
\item $\delta Q=0$. 
Note that in the limit $Q \rightarrow 0 $ one can reach $f(\lambda)>0$, for which black hole could not be overspun, thereby verifying the validity of the WCCC for black hole having a single rotation. This verifies the recently obtained result  Ref.~\cite{Shaymatov19a} that WCCC is obeyed for single rotation even at linear order accretion. Consider the numerical example: For $Q=0.5$, $J_{\psi}=0.322011$, $\delta J_{\psi}=0.001$, and $\alpha=0.01$ with $\lambda=0.1$ we get $f(\lambda)=0.000041 >0$. Thus WCCC would always hold good for neutral particle.  

\item $\delta J_{\psi}=0$.
It is well known that a four dimensional charged black hole could be overcharged \cite{Revelar-Vega17}. To be a bit more quantitative let's reconsider Eq.~(\ref{Eq:linear-order1}), for $Q=0.5$, $J_{\psi}=0.322011$, $\delta Q=0.003$, and $\alpha=0.01$ with $\lambda=0.1$, we get $f(\lambda)=-0.00048 <0$. With this we again verify the result of Ref.~\cite{Revelar-Vega17} that WCCC could as in four dimension be violated.

\end{itemize} 

Thus a five dimensional black hole with single rotation could be overcharged but not overspun. The natural question then arises what happens to five dimensional charged black hole with a single rotation -- could it be overcharged or overspun under bombardment of over charged particles?

\begin{itemize}

\item   {We know that black hole cannot be overspun but it could be over charged. When both charge and rotation are present, the outcome should depend on which one is greater than the other. The question is, does this dominance refer to black hole rotation and charge parameters or that of the impinging particles? It turns out that it refers to the parameters of the impinging particles. We will show this by numerical examples.} Let's begin with $\delta J_{\psi} < \delta Q$. The question is, what might happen in this case? To answer this question we must approach, as in previous ones, the problem quantitatively. For given  $Q=0.5$, $J_{\psi}=0.322011$, $\delta Q=0.003$, $\delta J_{\psi}=0.0001$, and $\alpha=0.01$ with $\lambda=0.1$ leads to $f(\lambda)=-0.0002445$, and so black hole could be over extremalized violating the CCC. {Let's now interchange black hole parameters and keep the rest of the parameters unchanged. That is,  $Q=0.353553$, $J_{\psi}=0.499394$, $\delta Q=0.003$, $\delta J_{\psi}=0.0001$, and $\alpha=0.01$ with $\lambda=0.1$, will give $f(\lambda)=-0.00001495 <0$, implying over extremalization.}             
  
\item  $\delta J_{\psi} > \delta Q$. Let's again consider the numerical exercise: Take {a) $Q=0.5$, $J_{\psi}=0.322011$ and b) $Q=0.353553$, $J_{\psi}=0.499394$ for given} $\delta Q=0.0003$, $\delta J_{\psi}=0.001$, and $\alpha=0.01$ with $\lambda=0.1$. That leads to a) $f(\lambda)=6.3832\times10^{-6}>0$ {and b) $f(\lambda)=50.1196\times10^{-6}>0$}. It cannot be over extremalized, and the WCCC continues to hold ground. 

{\item  $\delta J_{\psi} = \delta Q$. Let's  consider values of parameters as follows: { a) $Q=0.5$, $J_{\psi}=0.322011$ and b) $Q=0.353553$, $J_{\psi}=0.499394$ for given }$\delta Q=0.003$, $\delta J_{\psi}=0.003$, and $\alpha=0.01$ with $\lambda=0.1$, {we get a) $f(\lambda)=-0.000417867 <0$ and b) $f(\lambda)=-0.000116191 <0$}. This shows that black hole could reach over-extremal state when impinging particles have angular momentum equal to charge.  }
\end{itemize}

What emerges from this analysis is that black hole with single rotation for linear accretion obeys WCCC so long as $\delta Q < \delta J{\psi}$, and the opposite is true for $\delta Q \geq \delta J{\psi}$ {irrespective of relative dominance of black hole rotation and charge parameters}. In Fig.~\ref{fig2} we verify the above numerical analysis for $\delta Q > \delta J{\psi}$ and $\delta Q < \delta J{\psi}$, respectively.  Interestingly in the case of equality of angular momentum and charge of impinging particles, it is charge's interaction plays dominating role for over extremalizing process.

\begin{figure*}
\centering
  \includegraphics[width=0.45\textwidth]{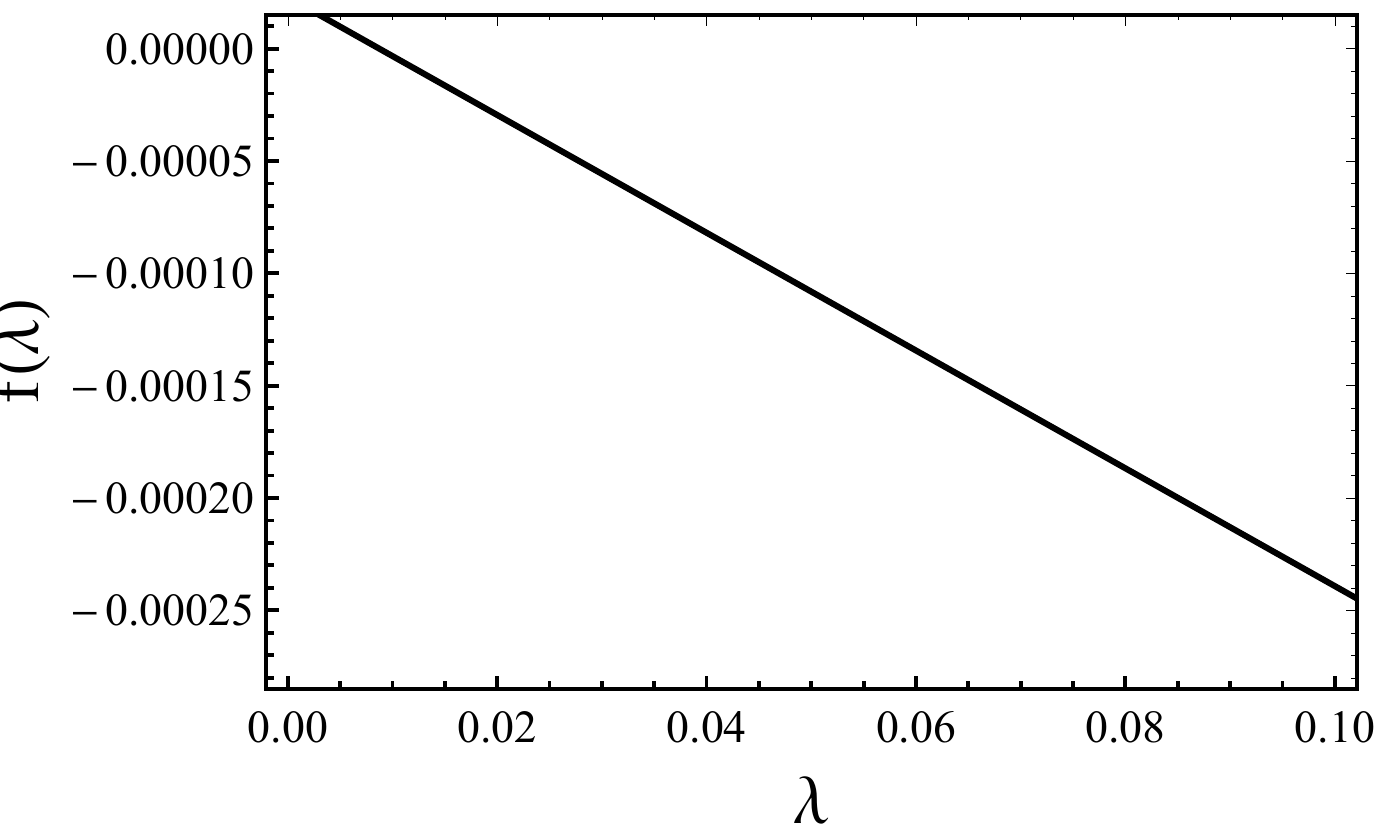}
  \includegraphics[width=0.45\textwidth]{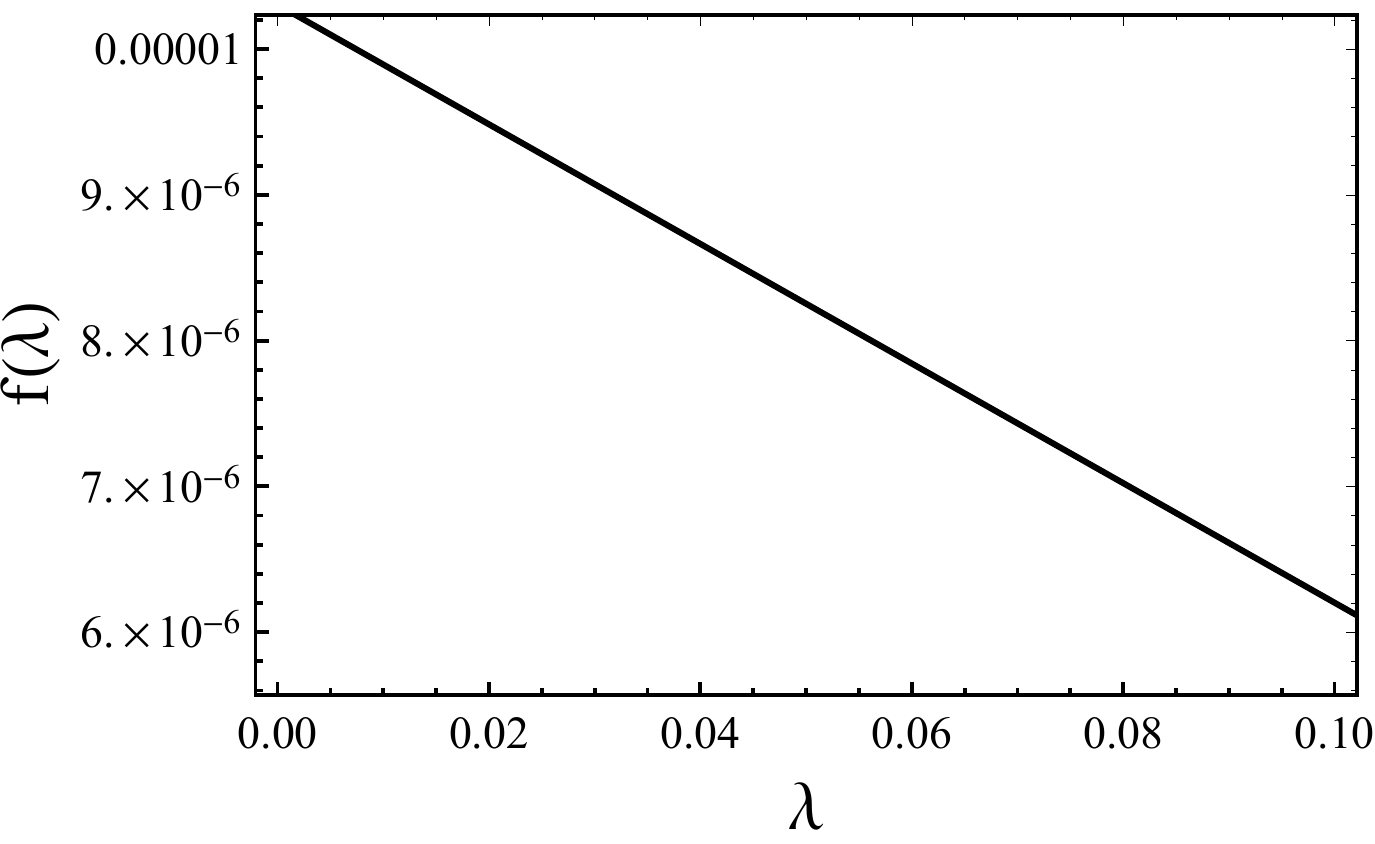}

\caption{\label{fig2} From left: $f(\lambda)$ for  $\delta J_{\psi} \ll\delta Q$ and $\delta J_{\psi}\gg \delta Q $ against $\lambda$ for the given values of test particle and black hole parameters. }
\end{figure*}

 \subsubsection{Non-linear order accretion}
 
Let's rewrite Eq.~(\ref{Eq:both}) in the case of a single rotation,                  
\begin{widetext}
\begin{eqnarray} \label{Eq:non-linear-order}
f(\lambda)&=&\left(\alpha-\frac{48\times 3^{3/4}Q^{3/2}\left(2 M+\sqrt{3} Q\right)^3\bigg(3\pi J_{\psi}\delta J_{\psi}+4Q\left(2M+\sqrt{3}Q\right)\delta Q\bigg)}{\left(9\sqrt{3}\pi J_{\psi}^2+4Q\left(2 M+\sqrt{3} Q\right)^2\right)^2} \lambda\right)^2 +\mathcal O(\alpha^3,\alpha^2\lambda,\alpha\lambda^2,\lambda^{3})\, .
\end{eqnarray}
\end{widetext}
From this, it is clear that black hole cannot be over extremalized when second order perturbations,  $\mathcal O(\lambda^2)$, are taken in. For non-linear accretion WCCC thus always holds good.


\section{Conclusions}
\label{Sec:Conclusion}

It is known that there does not exist a true analogue of four dimensional Kerr-Newman rotating charged black hole in five dimension. On the other hand there exists an analogue of Kerr rotating black hole in five or higher dimensions \cite{Myers-Perry86}. Strangely electric charge cannot be injected onto rotating black hole. However there exists a very close cousin of Kerr-Newman black hole in minimally gauged supergravity solution of rotating and charged black hole \cite{Chong05}. To this black hole we have in this paper extended the analysis of over extremalization under linear and non-linear accretion process \cite{An18}. 

In general it turns out that as is the case in for all other cases, over extremalizing is possible for linear order while it gets miraculously reversed when non-linear perturbations are included. The five dimensional black hole in question thus falls in line with all other black holes that WCCC could be violated at linear order but it is always restored back at non-linear order accretion. However there is a subtle exception for rotating black hole in five dimension which has two rotation axes permitting two rotation parameters. 

Very recently, some of us \cite{Shaymatov19a} had demonstrated a remarkable property of a black hole with single rotation. Unlike four dimensional black hole, it cannot be overspun even at the linear order accretion while it could be overspun when both rotations are present. This property is however carried through for the five dimensional rotating charged black hole under study. A charged black hole could always be overcharged under linear accretion. In this case there are both rotations and charge present. Hence the question, when would it be over extremalized and when not? As expected it turns out that when rotation parameter of impinging particle is greater than its charge, over extremalizing is prohibited while the opposite is the case when charge is greater than or equal to rotation parameter. It is interesting that in the case of equality of rotation and charge parameters, it is the latter's contribution that dominates. In all this relative dominance of charge or rotation of black hole is however irrelevant. 

As pointed out in \cite{Shaymatov19a}, a black hole with single rotation in five dimension is a different entity like extremal black hole. The latter can never be over extremalized and interestingly so is the case for the former as well. It seems when black hole has the maximum number of rotations that are permitted in a given spacetime dimension, it can be overspun under linear order accretion while if it has less than the maximum allowed, it cannot be overspun. In four dimension maximum allowed parameter is one and that is why it can be overspun while in five dimensions maximum allowed are two. That is why it can perhaps only violate WCCC when both rotations are present but not for single rotation. 

It may be noted that for non-linear accretion we have neat analytical expression showing $f(\lambda) > 0$ indicating absence of over-extremalization. However for linear order perturbations we had to resort to numerical evaluation because calculations were too involved and complicated. For over extremalization, any specific example is good enough to show that it occurs while for its absence one has to show that that it is never possible. We do however consider optimal choice of parameters which would indicate that the result would hold good in general for any other choice of parameters. Most importantly it is the non-linear regime that has the final and determining say which has been established rigorously and analytically.

\section*{Acknowledgments}
BA and SS acknowledge Inter-University Centre for Astronomy and Astrophysics, Pune, India, and Goethe University, Frankfurt am Main, Germany, for warm hospitality. {ND wishes to acknowledge visits to Albert Einstein Institute, 
Golm and to Astronomical Institute, Tashkent supported by the Abdus Salam International Centre for Theoretical Physics, Trieste under the Grant 
No. OEA-NT-01.} This research is supported in part by Projects No. VA-FA-F-2-008 and No. MRB-AN-2019-29 of the Uzbekistan Ministry for Innovative Development and by the Abdus Salam International Centre for Theoretical Physics under the Grant No. OEA-NT-01.

\bibliographystyle{apsrev4-1}  
\bibliography{gravreferences}

 \end{document}